# Large Thermoelectric Power Factor in P-type Si (110)/[110] Ultra-Thin-Layers Compared to Differently Oriented Channels

Neophytos Neophytou and Hans Kosina

Institute for Microelectronics, TU Wien, Gußhausstraße 27-29/E360, A-1040 Wien, Austria

e-mail: {neophytou|kosina}@iue.tuwien.ac.at

## Abstract

Using atomistic electronic structure calculations and Boltzmann semi-classical transport we compute the thermoelectric power factor of ultra-thin-body p-type Si layers of thicknesses from $W$=3nm up to 10nm. We show that the power factor for channels in [110] transport orientation and strong (110) surface confinement largely outperforms all differently oriented channels by more than 2X. Furthermore, the power factor in this channel increases by ~40% with layer thickness reduction. This increase, together with the large confinement effective mass of the (110) surface, make this particular channel less affected by the detrimental effects of enhanced surface roughness scattering and distortion at the nanoscale. Our results, therefore, point towards the optimal geometrical features regarding orientation and length scale for power factor improvement in 2D thin-layers of zincblende semiconductors.

# I. Introduction

The ability of a material to convert heat into electricity is measured by the dimensionless thermoelectric (TE) figure of merit $ZT=\sigma S^2 T/(\kappa_e+\kappa_l)$, where $\sigma$ is the electrical conductivity, $S$ is the Seebeck coefficient, $\kappa_e$ is the electronic part and $\kappa_l$ is the lattice part of the thermal conductivity. Some of the best thermoelectric materials are based on rare earth or toxic elements and exhibit $ZT\sim 1$, which corresponds to low efficiencies of the order of ~10% of the Carnot efficiency [1, 2, 3]. Recent breakthrough experiments, however, have demonstrated that nanostructured and low-dimensional channels can offer large improvements in $ZT$ compared to the raw materials' values. Such effects have been observed for 1D nanowires (NWs) [4, 5], 2D thin-layer superlattices [6, 7, 8, 9], as well as materials with embedded nanostructures [10, 11]. More importantly, this has been achieved for common semiconductor materials such as Si, SiGe and InGaAs [4, 5, 8, 9].

Most of this improvement has been attributed to a remarkable reduction in the phonon thermal conductivity $\kappa_l$ because of enhanced phonon scattering on the boundaries of narrow features and disorder [4, 8, 12, 13, 14, 15]. Narrow feature sizes, on the other hand, will in general degrade the electrical conductivity and power factor $\sigma S^2$ as well. In order to achieve efficient thermoelectric devices, $\sigma S^2$ needs to be kept high by proper optimization of the interplay between $\sigma$ and $S$. As we showed in previous works, at the nanoscale the transport and surface orientations as well as the confinement length scale are degrees of freedom through which electronic properties can be optimized [16, 17]. The sensitivity of the electronic properties to geometric parameters is especially strong in p-type nanoscale channels [16, 18].

In this work, we calculate the room temperature thermoelectric power factor of p-type Si ultra-thin-body (UTB) layers for thicknesses from $W$=3nm up to 10nm. Such channels, but also 2D superlattices formed of thin layers of these dimensions are promising candidates for TE applications [6, 7, 8, 9, 19]. We employ atomistic electronic structure and Boltzmann transport calculations. Our analysis shows that the variations in the electronic structure of the UTB layers with confinement and orientation can provide ways for power factor optimization. We demonstrate that the power factor of the (110)/[110] p-type Si channel outperforms by more than 2X the power factor for all other surface/transport orientations. In addition, we show that the power factor in this channel improves as the layer width is reduced down to 3nm, an effect that can potentially offset the detrimental effect of enhanced surface roughness scattering (SRS) with feature scaling. Our results, therefore, offer power factor optimization routes for high performance, thin-layer thermoelectric devices as well as thin 2D superlattice thermoelectric devices. The mechanisms we describe originate from features of the heavy-hole (HH) valence band, which are common in all zincblende semiconductors, and we therefore expect that our results would be qualitatively valid for other such semiconductors as well.

## II. Approach

We couple the 20 orbital atomistic $sp^3d^5s^*$-spin-orbit-coupled (SO) tight-binding (TB) model [20] to linearized Boltzmann transport theory [21, 22, 23]. This TB model accurately describes the electronic structure and inherently includes the effects of quantum confinement and orientation. It is a compromise between computationally expensive *ab-initio*, and inexpensive but less accurate effective mass methods. The electrical conductivity $\sigma$ and the Seebeck coefficient $S$ follow from linearized Boltzmann theory as:

$$\sigma = q_0^2 \int_{E_V}^{\infty} dE \left( -\frac{\partial f_0}{\partial E} \right) \Xi(E), \tag{1a}$$

$$S = \frac{q_0 k_B}{\sigma} \int_{E_V}^{\infty} dE \left( -\frac{\partial f_0}{\partial E} \right) \Xi(E) \left( \frac{E - E_F}{k_B T} \right), \tag{1b}$$

where the transport distribution function $\Xi(E)$ is defined as [24]:

$$\Xi(E) = \frac{1}{W} \sum_{k_{x,y},n} v_n^2(k_x) \tau_n(k_{x,y}) \delta(E - E_n(k_{x,y}))$$
$$= \frac{1}{W} \sum_n v_{k_x,n}^2(E) \tau_n(E) g_{2D}^n(E). \quad (2)$$

Here $v_{k_x,n}(E) = \frac{1}{\hbar} \frac{\partial E_n}{\partial k_x}$ is the group velocity in the transport direction, $\tau_n(k_{x,y})$ is the momentum relaxation time of a carrier with in-plane wave number $k_{x,y}$ in subband $n$, $g_{2D}^n(E_n)$ is the density of states for a 2D subband, $E_V$ is the valence band edge, $W$ is the width of the channel, and $E_F$ is the Fermi level.

We use Fermi's Golden rule to extract the momentum relaxation rates. We include scattering due to elastic acoustic phonons (ADP), inelastic optical phonons (ODP), and surface roughness (SRS), and use the full energy dependence for the momentum relaxation times. For computational efficiency, we make the following approximations: i) Confinement of phonons is neglected, and dispersionless bulk phonons are assumed. Instead, enhanced deformation potential values $D_{ODP}^{holes} = 13.24 \times 10^{10}$ eV/m and $D_{ADP}^{holes} = 5.34$ eV are employed, as is common practice for nanostructures [19, 21, 25, 26]. Such treatment could only affect our results quantitatively [27]. Our purpose, however, is to provide qualitative insight and design directions. ii) Surface relaxation is neglected. iii) For SRS we assume a 2D exponential autocorrelation function for the roughness with $\Delta_{rms}$ = 0.48nm and $L_C$ = 1.3nm and derive the transition rate from the shift in the band edges $\Delta E_V / \Delta W$ with confinement. As discussed by Uchida *et al.* [28], this is the strongest contribution to SRS in channels of a few nanometers in thickness. All these approximations are commonly employed in numerical calculations. Although in certain cases they might be quite strong, it is believed that they affect the results only quantitatively. Qualitatively, our results are determined mostly by the geometry-dependent electronic structure, which is the main focus of this work. The method is an extension to 2D of what we describe in [23] for 1D nanostructures.

## III. Results and Discussion

Before we describe the results obtained using the proper atomistic bandstructures, it is useful to estimate how the electronic structure affects $\sigma$ and $S$ using the simplified parabolic band approximation. For this we assume $\tau_n(E) \propto W/g_{2D}^n(E)$ and $v_n(E) \propto \sqrt{\tilde{I}}$, where $\tilde{I}$ , and $m_\parallel^*$ is the transport effective mass. We substitute these into Eq. 1, and after performing the summation over the subbands in Eqn. 2 (assuming single subband):

$$\sigma \propto \int_{E_V}^{\infty} \tilde{I} \left( \frac{f(E-E_F)}{\partial E} \right) dE$$

$$= \frac{1}{m_\parallel^*} \tilde{I}$$

(3)

where $v_{inj}$ is the carrier injection velocity, and $\tilde{I}$ is a function of $\eta_F = E_V - E_F$, independent of bandstructure at first order, and *exponentially* increasing with decreasing $\eta_F$. Similarly, from Eq. 1b, the Seebeck coefficient $S$ can be shown to follow:

$$S \propto \frac{\int_{E_V}^{\infty} F(\eta_F)\left(\frac{E-E_F}{k_B T}\right)dE}{\int_{E_V}^{\infty} F(\eta_F)dE}$$

(4)

where $F(\eta_F) = \tilde{I}\left(\frac{-E_F)}{\partial E}\right)$ appears in the numerator and denominator. The energy dependence of $S$ is, therefore, at first order independent of bandstructure [29]. Its magnitude reduces *linearly* as the subband energy is closer to the Fermi level (smaller $\eta_F$), as expected. At a certain carrier concentration, $\eta_F$ will depend on the density of states (DOS) of the dispersion (or the DOS effective mass, $m_{DOS}$). A large DOS will result in larger $\eta_F$ and larger $S$, but it will exponentially decrease $\sigma$.

The power factor $\sigma S^2$, therefore, depends on $v_{inj}$ and $\eta_F$. In UTB layers at a certain carrier concentration these two quantities are geometry dependent. Figure 1a shows the atomistically calculated hole $v_{inj}$ and $\eta_F$ for UTB layers on (100), (110) and (112) confinement surfaces, and in [100], [110] and [111] transport orientations as a function of the UTB film thickness, $W$. A hole concentration of p=$10^{19}$/cm$^3$ is assumed (a value close to where the peak of $\sigma S^2$ appears, as we show below). Strong anisotropic behavior is observed with respect to both surface and transport orientations. The (110)/[110] and (112)/[111] channels provide the highest velocities, followed by the (110)/[100] channel, whereas the (100) surfaces and the (112)/[110] channel have the lowest velocities. As the thickness of the UTB is scaled down, the hole velocities increase, especially for the first two channels. The reasons behind this are related to their bandstructure, and particularly how the curvature of the heavy-hole band along these directions changes under confinement. We will briefly explain this behavior below. The change in the bandstructure with confinement and orientation also results in different density of states (DOS) for each channel, which changes $\eta_F$ as well. Figure 1b shows the $\eta_F$ for the three surfaces with respect to the UTB layer thickness, at the same hole concentration of p=$10^{19}$/cm$^3$ for all channels. At larger thicknesses, $\eta_F$ is very similar in all cases because the films are thick enough for the DOS to approach the bulk DOS in all cases. As the UTB layer thickness is reduced, the $\eta_F$ for the (110) surface decreases. The $\eta_F$ in the (112) surface layers remains almost unchanged, whereas in the case of the (100) surface layers, $\eta_F$ increases.

The behavior of the $\eta_F$ as a function of surface orientation and layer width originates for the bandstructure of the channels, and how this changes with confinement and orientation. We provide an elaborate discussion regarding the bandstructures of thin p-type layers as well as nanowires in Refs [16, 17, 30, 31] and we refer the reader to those works for details. Here, we only provide a brief discussion on how these changes will affect $\eta_F$. Figure 2a shows the DOS(E) for (110) surface channels of widths $W$=10nm and $W$=3nm. Both functions are shifted to the same origin for comparison purposes. The insets show the corresponding bandstructures with the arrows pointing to the [110] direction. As the width is reduced, the bandstructure in the [110] direction acquires a larger curvature. The bands become lighter, which justifies the velocity increase in Fig. 1a for the (110)/[110] channel. The lighter bands, however, also result in smaller DOS(E) for the thinner UTB layer as shown in Fig. 2a. At a constant carrier concentration, the reduction in the DOS(E) will reduce $\eta_F$ as the width of the channel is reduced. The Fermi level will shift closer to the band edge in order to keep the carrier concentration constant. The situation is reversed for the (100) surface channels. Figure 2b shows the DOS(E) for the (100) channels of widths $W$=10nm and $W$=3nm. The electronic structure of these channels does not change significantly with confinement as shown by the insets of Fig. 2b. This is also reflected by the constant carrier velocities with width of the (100)/[100] and (100)/[110] channels in Fig. 1a. The 3D DOS(E) in this case, however, increases with confinement after the normalization by the width $W$. Assuming a simple effective mass approximation, the DOS is proportional to $M/W$, where $M$ is the number of subbands and $W$ is the normalization width of the thin layer. As the width is reduced, the number of subbands $M$ decreases, usually linearly for the thicker layers such that the ratio $M/W$ remains constant. At some point, only a few or even only one subband participates in transport. Further reduction of the width of the UTB layer will not be linearly compensated by a reduction in $M$, and the ratio $M/W$ will increase following $\sim 1/W$ as $M$ approaches closer to 1. The DOS(E), therefore, increases with confinement. Since the carrier concentration under a simple effective mass approximation is given by

$$n_{3D} = \frac{M}{W} \int_{E_V}^{\infty} g_{2D}(E) f(E-E_F) dE, \qquad (5)$$

in order to keep the carrier concentration $n_{3D}$ constant, the energy integral has to be reduced, which is achieved when the distance of the subband edges $E_V$ from the Fermi level $\eta_F = E_V - E_F$ is increased. The $\eta_F$ then increases as observed in Fig. 1b.

The larger the DOS(E), therefore, the larger the $\eta_F$ at a certain carrier concentration. Figure 2c shows the DOS(E) for the $W$=3nm thin layers of (100), (112) and (110) surfaces. The DOS(E) of the (100) layer is the largest, followed by that of the (112) layer, whereas that of the (110) layer is the smallest. Indeed, this follows the order in which $\eta_F$ appears in Fig. 1b.

The $v_{inj}$ and $\eta_F$ will determine $\sigma$ and $S$ for the UTB layers. Figure 3a shows the phonon-limited conductivity of the UTB channels as a function of the layer thickness $W$ at $p=10^{19}/cm^3$. Clearly, the (110)/[110] channel is advantageous compared to the other channels considered at all thicknesses. It is especially superior at smaller thicknesses, for which the conductivity of this channel largely improves (by ~3X), whereas that of the other channels shows either only slight improvement, or even slight degradation. This advantage of [110] p-type channels under strong (110) confinement is also verified by recent mobility measurements [32, 33, 34, 35]. This difference in performance between the different orientations originates from the fact that the (110)/[110] channels not only have the highest carrier velocities, but at the same carrier concentration they have the smallest $\eta_F$. Both quantities benefit the conductivity. On the other hand, the ~3X increase in $\sigma$ with thickness scaling for this channel can be justified from Eq. 3 by comparing $\sigma$ in the $W$=3nm and 10nm UTB channels:

$$\frac{\sigma_{3nm}}{\sigma_{10nm}} \propto \frac{v^2_{inj-3nm}}{v^2_{inj-10nm}} \cdot \frac{\tilde{\mathfrak{I}}}{\mathfrak{I}_F)}$$

$$\approx \left(\frac{1.22}{0.94}\right)^2 e^{\Delta\eta_F/k_BT} = 3.11 \quad (6)$$

where $\Delta\eta_F = 0.16 eV$ (from Fig. 1b).

The dependence of the Seebeck coefficient on the layer thickness in Fig. 3b follows the dependence of $\eta_F$ on thickness, as also explained above in Eq. 4. At larger thicknesses, $S$ is very similar for all channels. As the thickness is reduced, $S$ decreases in the (110) layers by ~30%, slightly decreases in the (112) layers, and increases in the (100) layers.

Since the $\sigma$ depends exponentially on $\eta_F$ as shown in Eq. 3 and Eq. 6, the power factor is more controlled by the electrical conductivity. The power factor trend with $W$ in Fig. 3c is very similar to the conductivity trend in Fig. 3a. The (110)/[110] channel outperforms the rest of the channels by more than 2X in the entire range of the examined thicknesses, despite the fact that *S* decreases for that channel at smaller thicknesses. In fact, the power factor increases by ~40% as the layer thickness is decreased. The performance of all other channels is lower and very similar to each other.

We note here that the orientation dependence of the power factor in p-type UTB layers does not have a one-to-one correspondence to that of p-type NWs presented in Ref. [23]. For the NWs, we showed that the [111] direction performs better than the [110] direction because of the lighter subbands that improve conductivity and reduce $\eta_F$. In the case of UTB layers, however, although the [111] oriented channel has high velocities, $\eta_F$ remains larger because of the higher DOS (Fig. 2c), resulting in reduced conductivity and power factor.

The large performance advantage for the (110)/[110] channel compared to the other channels is attributed to its larger electrical conductivity $\sigma$. It is not only larger in this channel compared to the rest, but it additionally increases for thinner channel widths. In reality, however, $\sigma$ deteriorates in nanostructures due to enhanced SRS. In Fig. 4 we include SRS in the calculations for $\sigma S^2$ (dashed lines). We assume that the influence of SRS originates from the shift in the band edges of the channel dispersions [28]. The Inset of Fig. 4 shows the band edges for the (110), (112) and (100) surfaces with respect to the layer thickness. The band edge in (110) films is the one affected the least, whereas the band edge in (100) films is the one affected the most by layer thickness fluctuations. This is an indication of a heavy confinement effective mass for the (110) surface, and a light one for the (100) surface.

The power factor in Fig. 4 is reduced once SRS is considered (here only results for the (110) and (100) surfaces in [110] transport are shown). The reduction originates solely from the reduction in $\sigma$ because the Seebeck coefficient is at first order independent of scattering, and it only marginally increases with SRS [23]. In the case of the (110) surface, SRS affects the conductivity and in extent the power factor only slightly, because of the weak shift in the band edges with confinement. The effect of SRS is stronger for the (100) surface where the band edges are more sensitive to confinement.

The (110)/[110] channel, therefore, not only outperforms the other channels, but it can also provide larger immunity to SRS. Thus, it can be the ideal candidate for p-type UTB layer thermoelectrics, and possibly for 2D in-plane superlattice thermoelectric materials. Quantitatively, the strength of SRS is determined by the roughness height $\Delta_{rms}$, and might possibly be stronger once additional Coulomb related effects are considered [36]. The point, however, is that the (110)/[110] channel, with the larger confinement effective mass, and the underlying bandstructure mechanism that causes the conductivity and the power factor to increase with thickness reduction, can compensate the detrimental effects of SRS. This is particularly important, because small feature sizes and roughness are necessary in order to achieve a large reduction in $\kappa_l$ and enhance the *ZT* figure of merit. For this purpose, rough nanowires [4, 5], thin-layers [6, 7, 8, 9], and lately nanoporous materials [10, 14] are currently receiving large attention. Such approaches, however, often degrade the power factor as well. The confinement and orientation dependences we describe provide guidance into how to still achieve high power factors in such channels, necessary for enhanced thermoelectric performance, and how to partially compensate for detrimental roughness and distortion effects.

A comparison between the performance of the p-type (110)/[110] UTB layers presented here, and the p-type [110] NWs we present in Ref. [23], shows that the stronger confinement in [110] NWs could provide somewhat larger power factors at narrower diameters of 3nm. However, [110] NWs suffer more from SRS because they are also confined by the strongly affected (100) surface, rather than only the weakly affected (110) surface. Once SRS is considered, the performance of the two channels is very similar. 2D thin layers, however, could offer the advantage of being more easily scaled to industrial processes than 1D NWs. Besides, the thermal conductivity in 2D layers can be as low as the one achieved in NWs for such small feature sizes. In recent works it was shown both by experiments and simulations that 2D thin layers, nanoporous thin films of Si or SiGe [10, 14, 37], and 2D superlattices composed of Si layers/Ge nanodots [15] could have thermal conductivities close to or even below the amorphous limit. Furthermore, it was shown that in some of these structures the electron transport is much less disrupted. This means that proper power factor optimization as we suggest in this work, not only in thin films but also in the thin-film-based structures we mention above, could potentially provide high *ZT* values at room temperature as well, similar to what has been measured in NWs [4, 5].

The UTB orientation comparison as of now was limited to a fixed carrier concentration of $p=10^{19}/cm^3$. We show here, however, that the performance advantage of the (110)/[110] channel holds for different hole concentrations as well. In Fig. 5 we show the phonon-limited power factor as a function of the carrier concentration for three different channels, the (110)/[110] channel with (i) *W*=3nm and (ii) *W*=10nm, and (iii) the (112)/[111] channel with *W*=3nm. The rest of the channels have similar or lower power factors than the (112)/[111] channel, at least for carrier concentrations below $p=10^{20}/cm^3$. For the sake of clarity we do not show them here. The (110)/[110] channel has a higher power factor in the entire carrier concentration range. For these orientations, the thinner *W*=3nm channel has a higher $\sigma S^2$ up to concentrations of $p=10^{19}/cm^3$, whereas at higher concentrations it loses this advantage to the thicker layers of the same channel orientation. Note, however, that it has the highest maximum power factor compared to the other layers, which peaks around carrier concentrations $p=7 \times 10^{18}/cm^3$.

## V. Conclusion

In summary, we have calculated the thermoelectric coefficients ($\sigma$, *S*, $\sigma S^2$) for silicon p-type ultra-thin-body layers with channel thicknesses from *W*=3nm to *W*=10nm using atomistic electronic structure and Boltzmann transport calculations. We have investigated various transport and confinement orientations. We find that the (110)/[110] channel shows a significant performance advantage compared to all other channel orientations (by more than 2X), and in addition, the phonon-limited power factor in such channel increases by ~40% as the (110) confinement increases. Furthermore, the (110) surface shows stronger immunity to the detrimental effect of SRS because of a larger confinement effective mass. These factors make the (110)/[110] channel an ideal candidate for ultra-thin p-type thermoelectric channels. Quantitatively, this conclusion is relevant not only for Si, but for other zincblende p-type materials with similar valence band features as well. Our results could provide guidance into design optimization strategies for high power factor in low-dimensional and nanostructured thermoelectric devices, in which narrow feature sizes are necessary to reduce the phonon part of the thermal conductivity $\kappa_l$ to achieve enhanced *ZT* figure of merit.

*Acknowledgement:* This work was supported by the Austrian Climate and Energy Fund, contract No. 825467.

Figure 1:

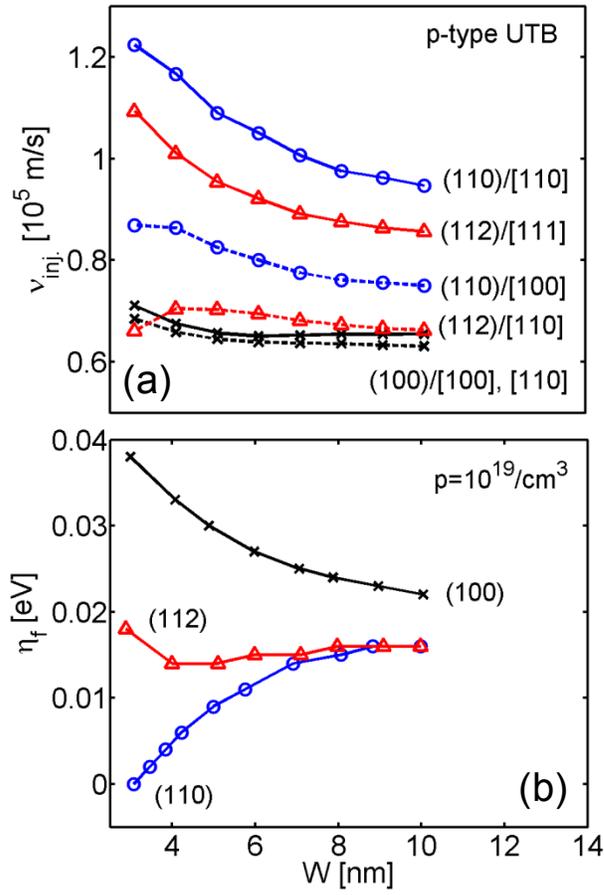

Figure 1 caption:

(a) The carrier injection velocity of the UTB layers vs. the layer thickness $W$. Channels of various confinement and transport orientations are shown. (b) The $\eta_F = E_V - E_F$ vs. $W$ for the various surfaces. Carrier concentration p=$10^{19}$/cm$^3$ is assumed for all cases.

Figure 2:

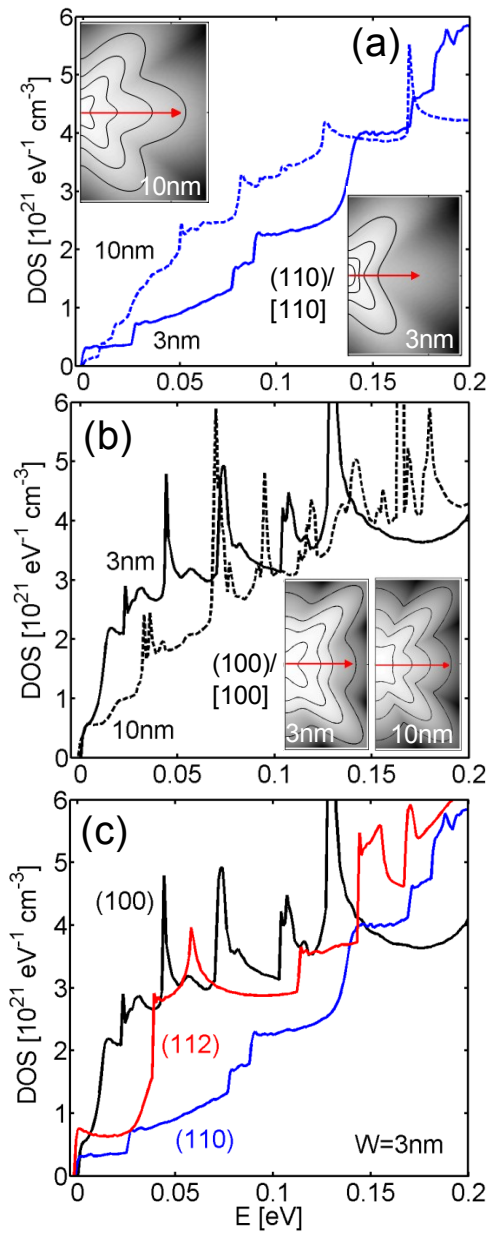

Figure 2 caption:

The DOS versus energy for UTB layers of different widths on different surfaces. (a) (110) surface and widths $W$=3nm (solid) and $W$=10nm (dashed). (b) (100) surface and widths $W$=3nm (solid) and $W$=10nm (dashed). Insets of (a) and (b): The corresponding $E(k)$ energy surfaces for the highest subbands. (c) The DOS(E) versus energy for the (100), (110), and (112) surfaces of the channels with $W$=3nm.

Figure 3:

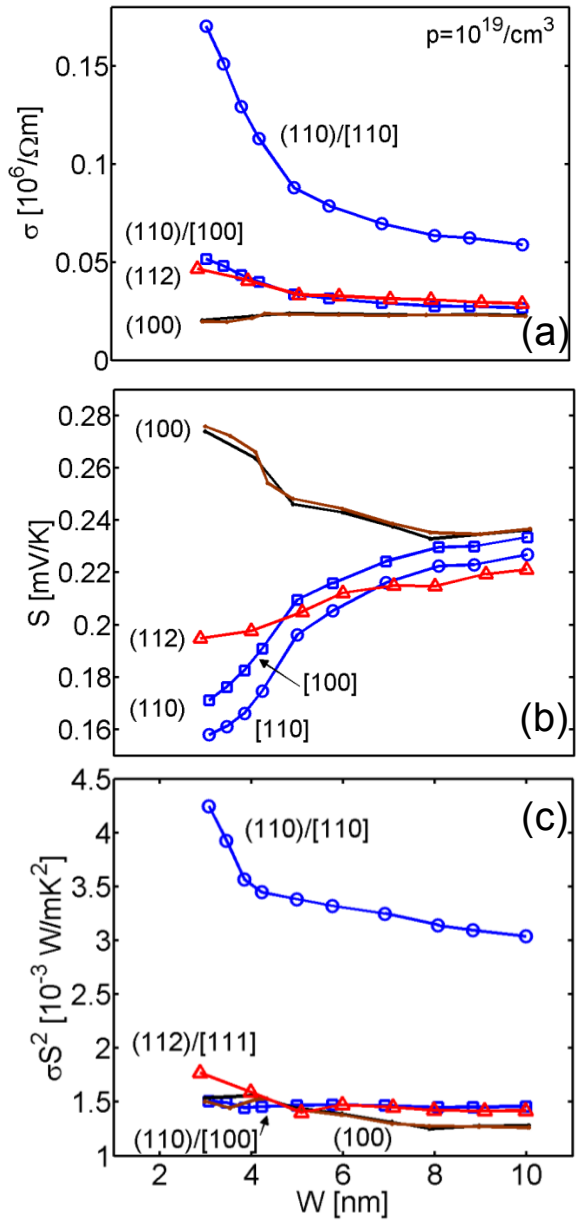

Figure 3 caption:

Phonon-limited thermoelectric coefficients for p-type UTB layers at $p=10^{19}/cm^3$ vs. layer thickness $W$: (a) electrical conductivity, (b) Seebeck coefficient, and (c) power factor. Various surface and transport orientations are presented as noted.

Figure 4:

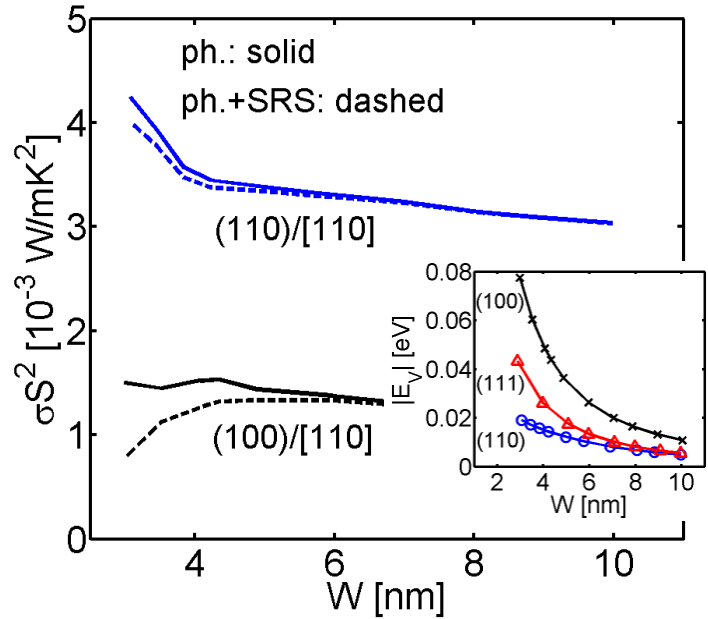

Figure 4 caption:

Power factor for p-type UTB layers at $p=10^{19}/cm^3$ vs. layer thickness $W$. [110] transport channels on (110) and (100) surfaces are presented as noted. Solid lines: Phonon-limited results. Dashed lines: Phonon plus SRS limited results. Inset: The valence band edges of the (100), (110), and (112) surfaces vs. $W$.

Figure 5:

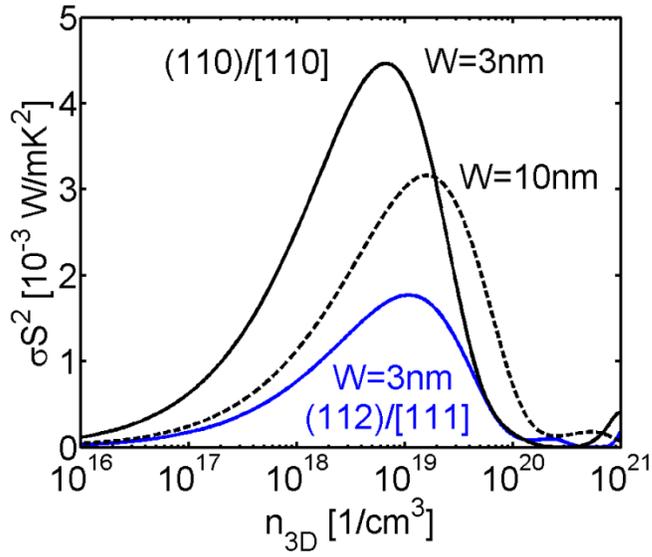

Figure 5 caption:

The phonon-limited thermoelectric power factor for UTB layers vs. the carrier concentration. UTB channels shown: i) (110)/[110] with $W$=3nm (solid-black). ii) (110)/[110] with $W$=12nmin (dashed-black), and iii) (112)/[111] with W=3nm (solid-blue).